\documentstyle [art12]{article}
\begin {document}
\bibliographystyle {plain}

\title
{\bf Skyrmion Liquid in SU(2)-invariant  Quantum Hall Systems}
\author{A. G. Green, I. I. Kogan  and A. M. Tsvelik}
\maketitle
\begin {verse}
Department of  Physics, University of Oxford, 1 Keble Road, Oxford OX1,  UK
\end{verse}
\begin{abstract}
We report on a theory of the Skyrmion states which occur in quantum
Hall regime near certain filling fractions.  It is shown that  in the
limit of zero Zeeman coupling in a
realistic temperature range the Skyrmion plasma  is a liquid described
by the effective model of massive two-dimensional Dirac fermions.



\end{abstract}


There are several physically different realizations of Quantum Hall
systems where electrons carry additional SU(2) degrees of freedom. In the
simplest case such degree of freedom is electron spin. In free space
electrons have the Landee factor $g_L = 2$ and the Zeeman splitting
$g_L\mu_B B$ is precisely equal to the distance beetween the Landau
levels $\omega_c$. However,  in real Quantum Hall devices
there are several reasons for the Zeeman
energy to be much smaller than the Landau splitting. The first reason
is the small effective mass ($m^* \sim 0.068 m_e$ in GaAs) which
leads to the increase in  the cyclotron frequency $\omega_c$, the second
is the spin orbit coupling which reduces $g_L$ by roughly a factor of
4. Thus the ratio $g_L\mu_BB/\omega_c$ in GaAs is about 0.02 which
makes spin fluctuations an important degree of freedom to be taken
into account. The Zeeman coupling may be further reduced by the
effects of pressure, perhaps even to zero.

A second example of a multi-component system is silicon where the
conduction band minimum occurs at six symmetry equivalent points lying
near the zone boundary. In the presence of the oxide barrier in a Si
MOSFET device and a very large confining electric field perpendicular to the
barrier the cubic symmetry is broken and only two valleys contribute
to the low energy properties. The remaining effective Hamiltonian has
the SU(2) symmetry just like a spin-1/2 system \cite{rasolt}.

Recently it has been proposed that multi-component Quantum Hall
systems may develop sophisticated spin (or pseudospin) textures when
the Landau level filling factor slightly deviates from
$\nu_0 = 1/(2m + 1)$\cite{sondhi}. The recent Knight shift
measurements give an experimental support to this prediction
\cite{barrett}. The incompressible
ground state of a two-dimensional electron gas at these filling
factors is ferromagnetic. The origin of ferromagnetism is due to
exchange energy; when Zeeman energy is weak or even absent
(as in Si devices) the direction of the
ferromagnetic order parameter described by the  unit vector
field  ${\bf n}(\tau, x)$ can deviate significantly from the direction of
the external magnetic field. According to
Refs. \cite{sondhi},\cite{fertig}, \cite{moon}, in incompressible Quantum Hall
states the
density of topological charge of the vector field ${\bf n}$ is directly
related to the density of the extra electric charge:
\begin{equation}
q = - \frac{\nu}{8\pi}\epsilon_{\mu\nu}\left({\bf
n}.[\partial_{\mu}{\bf n}\times \partial_{\nu}{\bf n}]\right)
\end{equation}
The same authors suggest the following Lagrangian to describe  spin textures:
\begin{eqnarray}
{L} &=& \int d^2x[{\cal L}_1 + {\cal L}_2],\\
{\cal L}_1 &=& \frac{\nu}{4\pi l_B^2} {\bf A}\left[{\bf n}({\bf r}
)\right]\partial_t {\bf n} ({\bf r} )
+\frac{\rho_s}{2}|\partial_{\mu}{\bf n}|^2  +\frac{2\mu q}{\nu},
\label{action}\\
{\cal L}_2 &=& \frac{1}{2} \int d^2x'q(r)V(r - r')\frac{e^2}{\epsilon |r -
r'|}q(r')
- g_LB n^z, \label{L2}
\end{eqnarray}
where ${\bf A}\left[{\bf n}({\bf r} )\right]$ is the vector potential of a
unit monopole in spin space, i.e., ${\bf \nabla}_{{\bf n}} \times {\bf A}={\bf
n}$.
At large distances $|r| >> l_B$ the interaction is via the
ordinary Coulomb potential: $V(r - r') = \frac{e^2}{\epsilon |r -
r'|}$. ${\cal L}_1$ is the Lagrangian density of a ferromagnet with a non-zero
topological charge stabilized by the chemical potential $\mu$. The
quantity $\rho_s$ is the stiffness; its physical origin is the loss of
exchange and correlation energy when the spin orientation varies in
space. The estimates made for the case when the exchange energy is of the
Coulomb origin  give \cite{moon}
\begin{eqnarray}
\rho_s &=& a\frac{e^2}{\epsilon l_B} \nonumber\\
a &=& 2.49\times 10^{-2} (\nu = 1),  \: 9.23\times 10^{-4} (\nu = 1/3),
\: 2.34\times 10^{-4} (\nu = 1/5)
\end{eqnarray}
where $l_B$ is the magnetic length.
For $\nu = 1$ the stiffness $\rho_s \approx 4K$ for $B = $10T.

Let us start from the case when the Coulomb interaction and the
Zeeman energy are absent (the former one will be reintroduced later).
We choose the following  convinient
parametrization of the vector field, ${\bf n}$:
\begin{eqnarray}
w = \frac{n_x + \mbox{i}n_y}{1 - n^z}, \: n_x + \mbox{i}n_y =
\frac{2w}{1 + |w|^2}, \: n^z = \frac{|w|^2 - 1}{1 + |w|^2}
\end{eqnarray}
In these notations the static part of the Lagrangian
(\ref{action}) is,
\begin{equation}
L_1 = \int \frac{4d^2x}{(1 +
|w|^2)^2}\left[\rho_s(|\partial w|^2 +
|\bar\partial w|^2)  -
\frac{\mu}{4\pi}(|\partial w|^2 -
|\bar\partial w|^2)\right] \label{energy}
\end{equation}
where $z = x +iy$, $\partial=\partial/\partial z$ and
$\bar\partial=\partial/\partial{\bar z}$.

The energy is minimised on configurations called instantons or
Skyrmions \cite{belavin}:
\begin{equation}
w(z) = h\prod_{i=1}^{N}\left(\frac{z - a_i}{z - b_i}\right) \label{w}
\end{equation}
where $a,b,h$ are parameters, $a_i$ and $ b_i$ being called
coordinates of the instanton. According to ref.\cite{belavin} the
energy of such configuration is proportional to its topological
charge and does not depend on $\{h, a_i,b_i\}$: $E = (4\pi\rho_s - \mu)N$.

To find the partition function
\begin{equation}
Z = \int D{{\bf n}}\delta({{\bf n}}^2(x) - 1)\exp[ - \frac{1}{T}(E - \mu N)]
\end{equation}
in a semiclassical approximation it is necessary to expand the energy
functional (\ref{energy}) near the classical solutions (\ref{w}) and
after calculating the  corresponding
Gaussian integral over the fluctuations sum the contributions with
different values of parameters $h, a_i, b_i$. Such a program was
carried out by Fateev {\it et al.} \cite{fateev} (see also
\cite{polyakov}) who obtained the
following result:
\begin{equation}
Z = \sum_N\frac{1}{(N!)^2}\left(\frac{m}{2\pi}\right)^{2N}\int
\frac{d^2h}{(1 + |h|^2)^2}\prod_{i = 1}^N d^2a_i d^2b_i\frac{\prod_{i < j}|a_i
- a_j|^2|b_i -
b_j|^2}{\prod_{i,j}|a_i - b_j|^2}
\end{equation}
where
\begin{equation}
m \approx l_B^{-1}\left(\frac{T}{\rho_s-\mu/4\pi}\right)\exp[-
\frac{2\pi(\rho_s - \mu/4\pi)}{T}]
\end{equation}
This equation shows that the grand partition function of Skyrmions
coincides  with the partition function for the classical
two-dimensional Coulomb plasma at temperature $t = 1/4\pi$ with $a_i$
and $b_i$ being coordinates of positive and negative charges. Such
partition function can also be written as the partition function of
the free Dirac fermions:
\begin{equation}
Z = \int D\psi D\bar\psi \exp[ \int
d^2x\bar\psi(\mbox{i}\gamma_{\alpha}\partial_{\alpha} - m)\psi]
\end{equation}
In order to express $m$ in terms of the extra charge density of the
Quantum Hall state, we have to differentiate the free energy with
respect to $\mu$:
\begin{equation}
q \equiv \frac{1}{2\pi l_B^2}\left|\frac{\nu}{\nu_0} - 1\right| = -
\frac{\partial\Omega}{\partial\mu} = T\frac{\partial
m^2}{\partial\mu}\int \frac{d^2k}{(2\pi)^2}\frac{1}{k^2 + m^2} =
\frac{m^2}{\pi}\ln\left(1/m\lambda\right) \label{q}
\end{equation}
where $\lambda \sim \frac{e^2}{\epsilon T}$ is the minimal distance
between the Skyrmions determined by the Coulomb repulsion
(renormalization of the minimal distance cut-off is the only plausable
effect of the Coulomb interaction one may think of). Solving Eq.(\ref{q})
with respect to $m$ we obtain with the logarithmic accuracy
\begin{equation}
m^2 \approx - 2\pi q/\ln(\pi\lambda^2 q) \label{m}
\end{equation}

Let us discuss conditions which must be fulfilled for this result to
be self-consistent. Firstly we have considered only static
configurations of the field. This is equivalent to projecting out the
zeroth Matsubara frequency and so we introduce an energy cut off at the
first Matsubara frequency, $\omega = 2\pi T$. The spectrum of magnons
is given by \cite{moon} $\omega = \frac{4\pi\rho_s}{\nu}(kl_B)^2$
which implies  a momentum cut off
$\Lambda = \frac{1}{l_B}\left(\frac{ T}{2\rho_s}\right)^{\frac{1}{2}}$. In
the absence of s
Skyrmions the static O(3) non-linear sigma model decays
exponentially at distances larger than,
\begin{equation}
\xi \sim \Lambda^{-1} \frac{T}{\rho_s}\exp\left(2\pi\frac{\rho_s}{T}\right)
\sim
l_B\left(\frac{T}{\rho_s}\right)^{\frac{1}{2}}\exp\left(2\pi\frac{\rho_s}{T}\right)
\end{equation}
If $q\xi^2 < 1$
the instantons are screened by fluctuations around the classical
minima, interact weakly  and their influence is small. Thus we need
$q\xi^2 >> 1$, that is,
\begin{equation}
T << 2\pi\rho_s \label{cond1}
\end{equation}
Secondly, we have neglected quantum
fluctuations. The magnon spectrum means that these fluctuations may be
neglected if
\begin{equation}
2\pi T >> \frac{4\pi\rho_s}{\nu}(ml_B)^2
\end{equation}
which, using equations (\ref{q}) and (\ref{m}), leads to the condition
\begin{equation}
T >> T_c \equiv \frac{2\rho_s}{\nu}\left|\frac{\nu}{\nu_0} -
1\right|\left\{\ln\left(\frac{2l_B^2}{\lambda^2|\nu/\nu_0 -
1|}\right)\right\}^{-1}\label{cond2}
\end{equation}
Finally, the Coulomb repulsion means that we must have $k\lambda^2<<1$
which gives the further constraint
\begin{equation}
T>>m^{\frac{1}{2}}\frac{e^2}{\epsilon} \sim
\left|\frac{\nu}{\nu_0}-1\right|^{\frac{1}{4}}\frac{e^2}{\epsilon l_B^2}
\end{equation}
Note that the length $m^{-1}$ is much smaller than $\xi$ and that
it has a much slower temperature dependence; logarithmic compared with
exponential.

Both conditions (\ref{cond1}) and (\ref{cond2}) are entirely
realistic. In this temperature range the system is in a liquid phase
with a finite correlation length $\sim m^{-1}$ with $m$ given by Eq.(\ref{m}).

As we have remarked above, in the absence of the Zeeman energy the
Coulomb interaction just prevents Skyrmions from approaching each
other too closely. If $|\nu/\nu_0 - 1|\lambda^2 << 1$ the Coulomb
interaction just renormalizes the ultraviolet cut-off. Therefore we
have solved the problem for systems with zero Landee factor. In the
realistic temperature range these systems are in a liquid phase. The
case of finite Zeeman coupling will be presented elsewhere.

 We are grateful to Boris Altshuler,
John Chalker and Robin Nicolas for their interest in
the work.

\end{document}